\documentstyle[epsf,preprint]{ptptex}

\newcommand{\bra}[1]{\langle {#1} |}     
\newcommand{\ket}[1]{| {#1} \rangle}     
\newcommand{\wtilde}[1]{\widetilde{#1}} 

\title{
A Note on the Eigenvalue Problem\\
in the $su(1,1)$-Algebra}

\author{
Seiya {\sc Nishiyama}$^{1}$, 
Constan\c{c}a {\sc Provid\^encia},$^{2}$\\
Jo\~ao da {\sc Provid\^encia}$^{2}$, Yasuhiko {\sc Tsue}$^{1}$ 
and Masatoshi {\sc Yamamura}$^{3}$
}

\inst{
$^{1}$Physics Division, Faculty of Science, Kochi University, Kochi 780-8520, 
Japan\\
$^{2}$Departamento de Fisica, Universidade de Coimbra, P-3000 Coimbra, 
Portugal\\
$^{3}$Faculty of Engineering, Kansai University, Suita 564-8680, Japan
}

\recdate{
\today
}

\abst{
Normalization constant in the eigenstate appearing in the eigenvalue problem 
of the $su(1,1)$-algebra is discussed. 
This normalization constant is expressed in terms of the Gauss' hypergeometric 
series which is not absolutely convergent. 
It is proved that this series is obtained as a certain limit of 
an absolutely convergent series, which 
was conjectured in the previous paper. 
}

\begin{document}

\maketitle


In the previous paper, hereafter referred to as (A),\cite{1} 
we investigated the eigenvalue problem for the $su(1,1)$-algebra. 
Concerning this problem, we encountered a certain trouble for the normalization 
constant of the eigenstate. 
This normalization constant is expressed in terms of the Gauss' hypergeometric 
series which is not absolutely convergent. 
Therefore, it is impossible to fix the normalization constant. 
In (A), in order to escape from this trouble, we gave a conjecture. 
The main aim of this paper is to clarify this conjecture 
in the framework which is more extensive than that in (A).

First, we introduce the following operator : 
\begin{subequations}\label{1}
\begin{eqnarray}
& &{\hat T}_0(\theta)=\cos ^2\theta\cdot {\hat T}_0+\sin^2\theta\cdot (1/2)
({\hat T}_++{\hat T}_-) \ , 
\label{1a}\\
& &{\hat T}_0(\theta)^*={\hat T}_0(\theta) \ . 
\label{1b}
\end{eqnarray}
\end{subequations}
The set $({\hat T}_{\pm,0})$ obeys the $su(1,1)$-algebra and its properties 
are listed in the relations (A$\cdot$2$\cdot$1) and (A$\cdot$2$\cdot$2). 
If $\theta=0$ and $\pm\pi/2$, ${\hat T}_0(\theta)$ is reduced to 
\begin{subequations}\label{2}
\begin{eqnarray}
& &{\hat T}_0(0)= {\hat T}_0 \ , 
\label{2a}\\
& &{\hat T}_0(\pm\pi/2)=(1/2)\cdot({\hat T}_++{\hat T}_-) \ . 
\label{2b}
\end{eqnarray}
\end{subequations}
In (A), also in Ref.\citen{2}, the eigenvalue problem of the $su(1,1)$-algebra 
was investigated for the operators $({\hat {\mib T}}^2,{\hat T}_0)$ and 
$({\hat {\mib T}}^2,(1/2)({\hat T}_++{\hat T}_-))$. 
Here, ${\hat {\mib T}}^2$ denotes the Casimir operator defined in the 
relation (A$\cdot$2$\cdot$3). 
Therefore, ${\hat T}_0(\theta)$ is in the intermediate position of the 
above three cases. In associating with ${\hat T}_0(\theta)$, we introduce the 
following operators : 
\begin{subequations}\label{3}
\begin{eqnarray}
& &{\hat T}_\pm(\theta)= A_{\pm}(\theta)\cdot{\hat T}_+ + 
B_{\pm}(\theta)\cdot{\hat T}_- + 
C_{\pm}(\theta)\cdot{\hat T}_0 \ , 
\label{3a}\\
& &{\hat T}_\pm(\theta)^*\neq {\hat T}_{\mp}(\theta) \ . 
\label{3b}
\end{eqnarray}
\end{subequations}
In (A), the three cases were treated : 
\begin{subequations}\label{4}
\begin{eqnarray}
& &{\hat T}_\pm(0)= {\hat T}_\pm \ , 
\label{4a}\\
& &{\hat T}_\pm(+\pi/2)=(1/2i)\cdot({\hat T}_+ -{\hat T}_-)\mp{\hat T}_0 \ , 
\nonumber\\
& & {\hat T}_\pm(-\pi/2)=(1/2i)\cdot({\hat T}_+ -{\hat T}_-)\pm{\hat T}_0 \ .
\label{4b}
\end{eqnarray}
\end{subequations}
We require that $A_{\pm}(\theta)$, $B_{\pm}(\theta)$ and $C_{\pm}(\theta)$ 
should be determined as functions of $\theta$ under the condition 
\begin{subequations}\label{5}
\begin{eqnarray}
& &[\ {\hat T}_+(\theta)\ , \ {\hat T}_-(\theta)\ ]=-2\lambda_0(\theta)\cdot
{\hat T}_0(\theta) \ , 
\label{5a}\\
& &[\ {\hat T}_0(\theta)\ , \ {\hat T}_\pm(\theta)\ ]
=\pm\lambda_\pm(\theta)\cdot{\hat T}_\pm(\theta) \ . 
\label{5b}
\end{eqnarray}
\end{subequations}
Here, $\lambda_0(\theta)$ and $\lambda_\pm(\theta)$ should also be determined 
as functions of $\theta$. 
The definition (\ref{4}) gives us 
\begin{subequations}\label{6}
\begin{eqnarray}
& &\lambda_0(0)=1 \ , \qquad \lambda_\pm(0)=1 \ , 
\label{6a}\\
& &\lambda_0(+\pi/2)=-i \ , \qquad \lambda_\pm(+\pi/2)=+i \ , \nonumber\\
& &\lambda_0(-\pi/2)=+i \ , \qquad \lambda_\pm(-\pi/2)=-i \ . 
\label{6b}
\end{eqnarray}
\end{subequations}
By substituting the forms (\ref{1}) and (\ref{3}) into the relation 
(\ref{5}), we have 
\begin{subequations}\label{7}
\begin{eqnarray}
& &A_+B_- - A_-B_+ = \lambda_0 \cos^2\theta  \ , \nonumber\\
& &A_+C_- - A_-C_+ = \lambda_0 \sin^2\theta  \ , \nonumber\\
& &B_+C_- - B_-C_+ = -\lambda_0 \sin^2\theta  \ , 
\label{7a}\\
& &A_\pm \cos^2\theta -(1/2)C_{\pm}\sin^2\theta = \pm \lambda_{\pm}A_{\pm} \ , 
\nonumber\\
& &B_\pm \cos^2\theta -(1/2)C_{\pm}\sin^2\theta = \mp \lambda_{\pm}B_{\pm} \ , 
\nonumber\\
& &(A_{\pm}-B_{\pm})\sin^2\theta = \pm\lambda_{\pm}C_{\pm} \ . 
\label{7b}
\end{eqnarray}
\end{subequations}
The relations (\ref{7a}) and (\ref{7b}) give us the following results : 
\begin{subequations}\label{8}
\begin{eqnarray}
& &\lambda_+(\theta)=\lambda_-(\theta)=\lambda(\theta) \ , \qquad
\lambda(\theta)^2=\cos^2\theta-\sin^2\theta \ , 
\nonumber\\
& &\lambda_0(\theta)=\lambda(\theta)\cdot\gamma_+(\theta)\gamma_-(\theta) \ , 
\label{8a}\\
& &A_{\pm}(\theta)
=\frac{1}{2}\cdot\frac{\sin^4\theta}{\cos^2\theta\mp\lambda(\theta)}
\cdot\gamma_{\pm}(\theta) \ , \nonumber\\
& &B_{\pm}(\theta)
=\frac{1}{2}\cdot\frac{\sin^4\theta}{\cos^2\theta\pm\lambda(\theta)}
\cdot\gamma_{\pm}(\theta) \ , \nonumber\\
& &C_{\pm}(\theta)=\sin^2\theta\cdot \gamma_{\pm}(\theta) \ . 
\label{8b}
\end{eqnarray}
\end{subequations}

In the framework (\ref{5}), it is impossible to determine 
$\gamma_{\pm}(\theta)$ and to fix the sign of $\lambda(\theta)$ $(=
\pm\sqrt{\cos^2\theta-\sin^2\theta})$. 
This trouble may be solved by adopting an idea in which the results obtained 
in (A) are reproduced at $\theta=0$ and $\pm\pi/2$. 
This means the following conditions : 
\begin{subequations}\label{9}
\begin{eqnarray}
& &A_+(0)=B_-(0)=1\ , \qquad A_-(0)=B_+(0)=C_{\pm}(0)=0 \ , \nonumber\\
& &\lambda_0(0)=1 \ , \qquad \lambda(0)=1 \ , 
\label{9a}\\
& &A_\pm(+\pi/2)=1/2i \ , \qquad 
B_\pm(+\pi/2)=-1/2i \ , \qquad C_{\pm}(+\pi/2)=\mp 1 \ , \nonumber\\
& &\lambda_0(+\pi/2)=-i \ , \qquad \lambda(+\pi/2)=i \ , \nonumber\\
& &A_\pm(-\pi/2)=1/2i \ , \qquad 
B_\pm(-\pi/2)=-1/2i \ , \qquad C_{\pm}(-\pi/2)=\pm 1 \ , \nonumber\\
& &\lambda_0(-\pi/2)=i \ , \qquad \lambda(-\pi/2)=-i \ . 
\label{9b}
\end{eqnarray}
\end{subequations}
We note the form of $\lambda(\theta)^2$ shown in the relation (\ref{8a}). 
In the region $|\theta| \leq \pi/4$, $\lambda(\theta)^2 \geq 0$ and 
we get $\lambda(\theta)=\pm\sqrt{\cos^2\theta -\sin^2\theta}$ which are real. 
However, since $\lambda(0)=1$, we adopt the part of plus sign of 
$\lambda(\theta)$, i.e., $\lambda(\theta)=\sqrt{\cos^2\theta -\sin^2\theta}$. 
In the region $\pi/4<|\theta|\leq \pi/2$, 
$\lambda(\theta)^2<0$ and we have 
$\lambda(\theta)=\pm i\sqrt{\sin^2\theta -\cos^2\theta}$ which are pure 
imaginary. However, $\lambda(+\pi/2)=i$ and $\lambda(-\pi/2)=-i$, and then, 
we adopt $\lambda(\theta)=i\sqrt{\sin^2\theta -\cos^2\theta}$ for 
$\pi/4<\theta\leq \pi/2$ and 
$\lambda(\theta)=-i\sqrt{\sin^2\theta -\cos^2\theta}$ for 
$-\pi/2 \leq \theta <\pi/4$. 
The above consideration is summarized as follows : 
\begin{equation}\label{10}
\lambda(\theta)=s(\theta)\sqrt{\cos^2\theta -\sin^2\theta} \ . 
\end{equation}
Here, $s(\theta)$ is the step function defined as 
\begin{equation}\label{11}
s(\theta)=\cases{ -1 \ , \qquad (-\pi/2\leq \theta < -\pi/4) \cr
+1 \ , \qquad (-\pi/4\leq \theta \leq \pi/2) }
\end{equation}
Further, the relations (\ref{8}) and (\ref{9}) give us 
\begin{equation}\label{12}
\gamma_\pm(0)=1 \ , \qquad \gamma_{\pm}(+\pi/2)= \mp 1 \ , \qquad
\gamma_{\pm}(-\pi/2)=\pm 1 \ . 
\end{equation}
A possible choice of $\gamma_{\pm}(\theta)$ satisfying the relation (\ref{12}) 
is as follows : 
\begin{equation}\label{13}
\gamma_\pm(\theta)=\cos \theta \mp \sin \theta \ . 
\end{equation}
Thus, as a possible choice, we obtain the results 
\begin{eqnarray}
& &\lambda(\theta)=s(\theta)\sqrt{\cos^2\theta -\sin^2 \theta} \ , 
\nonumber\\
& &\lambda_0(\theta)=s(\theta)\left(\sqrt{\cos^2\theta -\sin^2 \theta}
\right)^3 \ , 
\label{14}\\
& &A_{\pm}(\theta)=\frac{1}{2}\cdot
\frac{\sin^4 \theta (\cos\theta\mp\sin\theta)}
{\cos^2\theta \mp s(\theta)\sqrt{\cos^2\theta-\sin^2\theta}} \ , 
\nonumber\\
& &B_{\pm}(\theta)=\frac{1}{2}\cdot
\frac{\sin^4 \theta (\cos\theta\mp\sin\theta)}
{\cos^2\theta \pm s(\theta)\sqrt{\cos^2\theta-\sin^2\theta}} \ , 
\nonumber\\
& &C_{\pm}(\theta)=\sin^2\theta (\cos \theta \mp \sin\theta) \ . 
\label{15}
\end{eqnarray}

Our next and the most important concern is to investigate the 
eigenvalue problem of the present system. 
It is governed by the $su(1,1)$-algebra, and then, first, we set up 
the following relations : 
\begin{eqnarray}
& &{\hat {\mib T}}^2\ket{t(\theta)}=t(t-1)\ket{t(\theta)} \ , 
\label{16}\\
& &{\hat T}_0(\theta)\ket{t(\theta)}=t(\theta)\ket{t(\theta)} \ , 
\label{17}\\
& &{\hat T}_-(\theta)\ket{t(\theta)}=0 \ . 
\label{18}
\end{eqnarray}
Here, we omit the extra quantum number. 
The relation (\ref{16}) supports that $\ket{t(\theta)}$ can be expanded in 
terms of the orthogonal set defined in the relation (A$\cdot$2$\cdot$9). 
We search $\ket{t(\theta)}$ in the form 
\begin{eqnarray}
& &\ket{t(\theta)}=\left(\sqrt{N_{t(\theta)}}\right)^{-1}
\exp(w(\theta){\hat T}_+)\ket{t} \ , 
\label{19}\\
& &{\hat T}_-\ket{t}=0 \ , \qquad {\hat T}_0\ket{t}=t\ket{t} \ , \qquad
{\hat {\mib T}}^2\ket{t}=t(t-1)\ket{t} \ . 
\label{20}
\end{eqnarray}
Here, $N_{t(\theta)}$ denotes normalization constant which will be discussed 
later. 
For obtaining $\ket{t(\theta)}$ explicitly, the following formula is useful : 
\begin{eqnarray}\label{21}
& &{\hat T}_0\ket{t(\theta)}=(t+w(\theta){\hat T}_+)\ket{t(\theta)} \ , 
\nonumber\\
& &{\hat T}_-\ket{t(\theta)}=(2w(\theta)t+w(\theta)^2{\hat T}_+)
\ket{t(\theta)} \ .
\end{eqnarray}
The definition of ${\hat T}_-(\theta)$ shown in the relation (\ref{3a}) and 
the use of the formula (\ref{21}) lead us to 
\begin{eqnarray}
& &2wB_- + C_- =0 \ , 
\label{22}\\
& &A_- + w^2B_- + w C_-=0 \ . 
\label{23}
\end{eqnarray}
The relation (\ref{22}) gives us 
\begin{equation}\label{24}
w(\theta)=-\frac{C_-(\theta)}{2B_-(\theta)}
=-\frac{\sin^2\theta}{\cos^2\theta +s(\theta)\sqrt{\cos^2\theta-\sin^2\theta}}
\ . 
\end{equation}
We can show that the form (\ref{24}) automatically satisfies the 
relation (\ref{23}). With the use of the result (\ref{24}) and the 
formula (\ref{21}), we can show that the state (\ref{19}) with 
(\ref{24}) satisfies the relation (\ref{17}) and $t(\theta)$ is given as 
\begin{equation}\label{25}
t(\theta)=\lambda(\theta)\cdot t=s(\theta)
\sqrt{\cos^2\theta-\sin^2\theta}\cdot t \ . 
\end{equation}
The relation (\ref{25}) denotes the eigenvalue of ${\hat T}_0(\theta)$ 
connecting to $t$ at $\theta=0$. 
We can learn that in the region $-\pi/4\leq\theta <\pi/4$, 
it is real and at the regions $-\pi/2\leq\theta <-\pi/4$ and 
$\pi/4 < \theta\leq\pi/2$, it is pure imaginary. 
The behavior is shown in Fig.1, which tells us that at the points 
$\theta=\pm\pi/4$, the phase change occurs.

\begin{figure}[hb]
  \epsfxsize=8cm  
  \centerline{\epsfbox{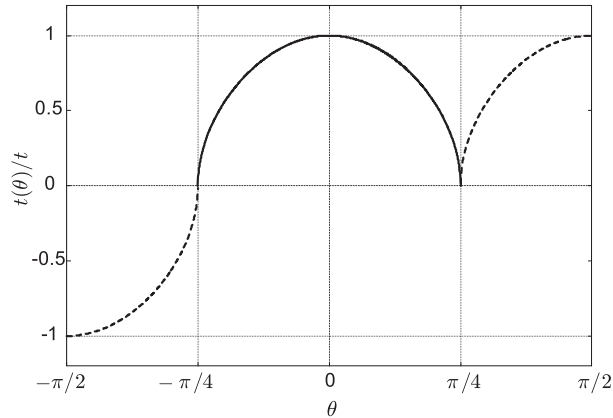}}
  \caption{The behavior of $t(\theta)/t$ is depicted as a function of
  $\theta$. The solid curve represents $t(\theta)/t$ which is real, 
  and the dotted curve 
  represent ${\rm Im} t(\theta)/t$ which is pure imaginary.}
   \label{fig1}
\end{figure}

In order to determine the normalization constant $N_{t(\theta)}$ shown 
in the ket state (\ref{19}), we must define the bra state $\bra{t(\theta)}$ 
which is conjugate to the ket state $\ket{t(\theta)}$. 
The following conditions for the bra state are required : 
\begin{eqnarray}
& &\bra{t(\theta)}{\hat {\mib T}}^2 = t(t-1)\bra{t(\theta)} \ , 
\label{26}\\
& &\bra{t(\theta)}{\hat T}_0(\theta) = t(\theta)\bra{t(\theta)} \ , 
\label{27}\\
& &\bra{t(\theta)}{\hat T}_+(\theta) = 0 \ . 
\label{28}
\end{eqnarray}
We can prove that the bra state $\bra{t(\theta)}$ satisfying the conditions 
(\ref{26})$\sim$(\ref{28}) is given in the form 
\begin{equation}\label{29}
\bra{t(\theta)}=\left(\sqrt{N_t(\theta)}\right)^{-1}\bra{t}
\exp(w(\theta){\hat T}_-) \ . 
\end{equation}
For the derivation of the form (\ref{29}), the following formula 
is useful : 
\begin{eqnarray}\label{30}
& &\bra{t(\theta)}{\hat T}_0=\bra{t(\theta)}(t+w(\theta){\hat T}_-) \ , 
\nonumber\\
& &\bra{t(\theta)}{\hat T}_+=\bra{t(\theta)}(2w(\theta)t
+w(\theta)^2{\hat T}_-) \ . 
\end{eqnarray}
Then, we define the normalization constant $N_{t(\theta)}$ as follows : 
\begin{eqnarray}
N_{t(\theta)}&=&\bra{t}\exp(w(\theta){\hat T}_-)\cdot
\exp(w(\theta){\hat T}_+)\ket{t} \nonumber\\
&=&\sum_{n=0}^{\infty}\frac{\Gamma(n+2t)}{n!\Gamma(2t)}(w(\theta)^2)^n \ , 
\label{31}\\
w(\theta)^2&=&
\frac{\cos^2\theta-s(\theta)\sqrt{\cos^2\theta-\sin^2\theta}}
{\cos^2\theta+s(\theta)\sqrt{\cos^2\theta-\sin^2\theta}} \ . 
\label{32}
\end{eqnarray}
The series (\ref{31}) is identical with the Gauss' hypergeometric series and 
it is absolutely convergent when $|w(\theta)^2|$ obeys 
\begin{equation}\label{33}
|w(\theta)^2| < 1 \ .
\end{equation}
Of course, the series (\ref{31}) is divergent when 
$|w(\theta)^2|>1$. Under the condition (\ref{33}), $N_{t(\theta)}$ is 
expressed in the form 
\begin{equation}\label{34}
N_{t(\theta)}=(1-w(\theta)^2)^{-2t}\ . \qquad (t>0)
\end{equation}
The behavior of $w(\theta)^2$ is divided into three groups : 
\begin{equation}\label{35}
w(\theta)^2=\cases{{\rm real} &\hbox{( $0\leq w(\theta)^2 <1 \ ; \ 
0\leq |\theta|<\pi/4$ )} \cr
{\rm real} &\hbox{( $ w(\theta)^2 =1 \ ; \ 
|\theta|=\pi/4$ )} \cr
{\rm imaginary} &\hbox{( $|w(\theta)^2|=1 \ ; \ 
\pi/4 < |\theta| \leq \pi/2$ )} }
\end{equation}
Therefore, in the region $0\leq |\theta|<\pi/4$, the form (\ref{34}) 
is available. However, the form (\ref{34}) cannot be used in the region 
$\pi/4 \leq |\theta| \leq \pi/2$. Then, as a conjecture, 
we suppose that the form (\ref{34}) is available even if in the region 
$\pi/4 \leq |\theta| \leq \pi/2$. This is in the same situation as 
that conjectured in (A). 
Later, we will give a reasonable interpretation for the conjecture. 
The behavior of $(1-w(\theta)^2)^{-1}$ is shown in Fig.2, in which we observe 
that except $|\theta|=\pi/4$, the state $\ket{t(\theta)}$ can exist and 
at the point $|\theta|=\pi/4$, 
the state $\ket{t(\pm\pi/4)}$ cannot be defined.

\begin{figure}[t]
  \epsfxsize=8cm  
  \centerline{\epsfbox{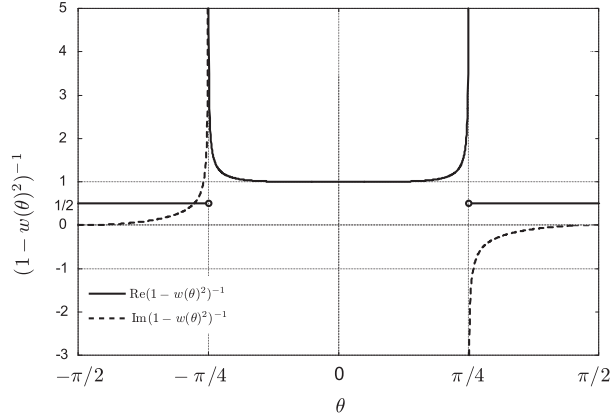}}
  \caption{The behavior of $(1-w(\theta)^2)^{-1}$ 
  is depicted as a function of
  $\theta$. The solid curve represents the ${\rm Re} (1-w(\theta)^2)^{-1}$ 
  and the dotted curves 
  represent ${\rm Im} (1-w(\theta)^2)^{-1}$.}
   \label{fig2}
\end{figure}

With the use of the relations (\ref{5}) and (\ref{16})$\sim$(\ref{18}), 
we can construct the ket state $\ket{t(\theta),t_0(\theta)}$ satisfying 
\begin{eqnarray}
& &{\hat {\mib T}}^2\ket{t(\theta),t_0(\theta)}
=t(t-1)\ket{t(\theta),t_0(\theta)} \ , 
\label{36}\\
& &{\hat T}_0(\theta)\ket{t(\theta),t_0(\theta)}
=t_0(\theta)\ket{t(\theta),t_0(\theta)} \ . 
\label{37}
\end{eqnarray}
The state $\ket{t(\theta),t_0(\theta)}$ is easily obtained in the form 
\begin{eqnarray}
& &\ket{t(\theta),t_0(\theta)}=\left(\sqrt{N_{t(\theta),t_0(\theta)}}
\right)^{-1}({\hat T}_+(\theta))^{t_0-t}\ket{t(\theta)} \ , 
\label{38}\\
& &t(\theta)=\lambda(\theta)\cdot t \ , \qquad t_0(\theta)=
\lambda(\theta)
\cdot t_0 \ , \nonumber\\
& &t_0=t,\ t+1,\ t+2,\cdots \ .
\label{39}
\end{eqnarray}
The proof is omitted. The normalization constant $N_{t(\theta),t_0(\theta)}$ 
is expressed as 
\begin{equation}\label{40}
N_{t(\theta),t_0(\theta)}=(\cos^2\theta-\sin^2\theta)^2
\frac{(t_0-t)!\Gamma(t_0+t)}{\Gamma(2t)} \ . 
\end{equation}
Thus, we obtain the eigenstate $\ket{t(\theta),t_0(\theta)}$.

We conjectured that the form (\ref{34}) is available even if 
$|w(\theta)^2|=1$, i.e., $\pi/4 < |\theta| \leq \pi/2$. 
Let us investigate the reason why it is permitted. For this purpose, 
we introduce the following operators in the region 
$\pi/4 < |\theta| \leq \pi/2$ : 
\begin{eqnarray}
& &{\wtilde T}_0(\theta)=\frac{\sqrt{1-\varepsilon}}{1-\varepsilon/2}
\left[{\hat T}_0(\theta)+\frac{1}{2}\frac{1}{\sqrt{\sin^2\theta-\cos^2\theta}}
\frac{i\varepsilon}{\sqrt{1-\varepsilon}}\cdot\frac{1}{2}
({\hat T}_+(\theta)-{\hat T}_-(\theta))\right] \ , \qquad
\label{41}\\
& &{\wtilde T}_\pm(\theta)=\frac{\sqrt{1-\varepsilon}}{1-\varepsilon/2}
\biggl[\frac{1}{2}\left(1+\frac{1-\varepsilon/2}{\sqrt{1-\varepsilon}}\right)
{\hat T}_\pm(\theta)-\frac{1}{2}\left(1-\frac{1-\varepsilon/2}
{\sqrt{1-\varepsilon}}\right){\hat T}_\mp(\theta) \nonumber\\
& &\qquad\qquad\qquad\qquad
\mp\frac{1}{2}\sqrt{\sin^2\theta-\cos^2\theta}
\frac{i\varepsilon}{\sqrt{1-\varepsilon}}{\hat T}_0(\theta))\biggl] \ . 
\label{42}
\end{eqnarray}
Here, $\varepsilon$ denotes a real parameter. The operators 
${\wtilde T}_{\pm,0}(\theta)$ obey 
\begin{equation}\label{43}
{\wtilde T}_0(\theta)^* \neq {\wtilde T}_0(\theta) \ , \qquad
{\wtilde T}_\pm(\theta)^* \neq {\wtilde T}_\mp(\theta) \ , \qquad
{\wtilde T}_\pm(\theta)^* \neq {\wtilde T}_\pm(\theta) \ .
\end{equation}
The important property is as follows : 
\begin{equation}\label{44}
{\wtilde T}_{\pm,0}(\theta) \longrightarrow {\hat T}_{\pm,0}(\theta) \ . 
\qquad (\varepsilon \rightarrow 0)
\end{equation}
The commutation relations are of the same forms as those of 
${\hat T}_{\pm,0}(\theta)$ shown in the relations (\ref{5a}) and (\ref{5b}) : 
\begin{subequations}\label{45}
\begin{eqnarray}
& &[\ {\wtilde T}_+(\theta)\ , \ {\wtilde T}_-(\theta)\ ]
=-2i s(\theta)\left(\sqrt{\sin^2\theta-\cos^2\theta}\right)^3
{\wtilde T}_0(\theta) \ , 
\label{45a}\\
& &[\ {\wtilde T}_0(\theta)\ , \ {\wtilde T}_\pm(\theta)\ ]
=\pm i s(\theta)\sqrt{\sin^2\theta-\cos^2\theta}\ {\wtilde T}_\pm(\theta) \ . 
\label{45b}
\end{eqnarray}
\end{subequations}
Here, we used the relation (\ref{14}) for $\lambda_0(\theta)$ and 
$\lambda(\theta)$. 
Then, our task is reduced to search the state $\ket{{\tilde t}(\theta)}$ 
obeying 
\begin{equation}\label{46}
{\wtilde T}_-(\theta)\ket{{\tilde t}(\theta)}=0 \ , \qquad
{\wtilde T}_0(\theta)\ket{{\tilde t}(\theta)}=
{\tilde t}(\theta)\ket{{\tilde t}(\theta)} \ . 
\end{equation}
Of course, this task may be performed in the framework of the exponential 
form 
\begin{equation}\label{47}
\ket{{\tilde t}(\theta)}=\left(\sqrt{N_{{\tilde t}(\theta)}}\right)^{-1}\exp(
{\wtilde w}(\theta){\hat T}_+)\ket{t} \ . 
\end{equation}
In order to avoid unnecessary complication and to respond directly to the 
promise given in (A), we show the result for the case $\theta=\pm\pi/2$. 
In these cases, ${\wtilde w}(\pm\pi/2)$ are given as 
\begin{equation}\label{48}
{\wtilde w}(\pm\pi/2)=\pm i\sqrt{1-\varepsilon} \ . 
\end{equation}
Therefore, we have 
\begin{equation}\label{49}
|{\wtilde w}(\pm\pi/2)^2|=1-\varepsilon \ . 
\end{equation}
The relation (\ref{49}) shows us 
\begin{equation}\label{50}
|{\wtilde w}(\pm\pi/2)^2|<1 \ . \qquad (\varepsilon >0)
\end{equation}
The normalization constant $N_{{\tilde t}(\pm\pi/2)}$ is given in the form 
\begin{equation}\label{51}
N_{{\tilde t}(\pm\pi/2)}=\sum_{n=0}^{\infty}\frac{\Gamma(n+2t)}{n!\Gamma(2t)}
\left({\wtilde w}(\pm\pi/2)^2\right)^{n} \ . 
\end{equation}
Then, for $\varepsilon>0$, the series (\ref{51}) is absolutely convergent 
and $N_{{\tilde t}(\pm\pi/2)}$ can be expressed as 
\begin{equation}\label{52}
N_{{\tilde t}(\pm\pi/2)}=\left(1-{\wtilde w}(\pm\pi/2)^2\right)^{-2t}
=(2-\varepsilon)^{-2t} \ .
\end{equation}
Clearly, the series (\ref{51}) is absolutely convergent and we have the form 
(\ref{52}). Therefore, we can define $N_{{t}(\pm\pi/2)}$ in terms of 
\begin{equation}\label{53}
N_{{t}(\pm\pi/2)}=\lim_{\varepsilon \rightarrow 0+}
N_{{\tilde t}(\pm\pi/2)} \ .
\end{equation}
The above is a possible reply of the promise mentioned in (A), i.e., 
the proof of the conjecture given in (A). 
The same conclusion as the above may be obtained in other region 
for $\theta$.

\acknowledgement

The main part of this paper was completed when the authors S.N., Y.T. and 
M.Y. stayed at Coimbra in September of 2004. 
They express their sincere thanks to Professor J. da Provid\^encia, 
co-author of this paper, for his kind invitation.


\begin{thebibliography}{99}
\bibitem{1}
S. Nishiyama, C. Provid\^encia, J. da Provid\^encia, Y. Tsue 
and M. Yamamura, submitted to Prog. Theor. Phys.
\bibitem{2}
E. Celeghini, M. Rasetti, M. Tarlini and G. Vittiello, Mod. Phys. Lett. 
{\bf B 3} (1989), 1213.\\
E. Celeghini, M. Rasetti and G. Vittiello, Ann. of Phys. {\bf 215} (1992), 156.
\end{thebibliography}
\end{document}